\documentclass[conference]{IEEEtran}
\IEEEoverridecommandlockouts
\usepackage{cite}
\usepackage{amsmath,amssymb,amsfonts}
\usepackage{algorithmic}
\usepackage{graphicx}
\usepackage{textcomp}
\usepackage{xcolor}
\usepackage{color, colortbl}

\usepackage{tabularx,booktabs}
\usepackage{rotating}
\newcommand{\RNum}[1]{\lowercase\expandafter{\romannumeral #1\relax}}

\newcolumntype{C}{>{\centering\arraybackslash}X} 
\usepackage{lipsum} 

\def\BibTeX{{\rm B\kern-.05em{\sc i\kern-.025em b}\kern-.08em
    T\kern-.1667em\lower.7ex\hbox{E}\kern-.125emX}}

\definecolor{LightCyan}{rgb}{0.75,1,1}

\IEEEoverridecommandlockouts\IEEEpubid{\makebox[\columnwidth]{ 979-8-3503-3439-5/23
/\$31.00~\copyright~2023 IEEE \hfill} \hspace{\columnsep}\makebox[\columnwidth]{ }}

\begin{document}

\title{
WEARS: Wearable Emotion AI with Real-time Sensor data

}

\author{\IEEEauthorblockN{Dhruv Limbani}
\IEEEauthorblockA{\textit{School of Computing} \\
\textit{SRMIST}\\
Kattankulathur, India \\
dl4267@srmist.edu.in}
\and
\IEEEauthorblockN{Daketi Yatin}
\IEEEauthorblockA{\textit{School of Computing} \\
\textit{SRMIST}\\
Kattankulathur, India \\
dy0787@srmist.edu.in}
\and
\IEEEauthorblockN{Nitish Chaturvedi}
\IEEEauthorblockA{\textit{School of Computing} \\
\textit{SRMIST}\\
Kattankulathur, India \\
nc7219@srmist.edu.in}
\and
\IEEEauthorblockN{Vaishnavi Moorthy}
\IEEEauthorblockA{\textit{School of Computing} \\
\textit{SRMIST}\\
Kattankulathur, India \\
vaishnam@srmist.edu.in}
\and
\IEEEauthorblockN{Pushpalatha M}
\IEEEauthorblockA{\textit{School of Computing} \\
\textit{SRMIST}\\
Kattankulathur, India \\
pushpalm@srmist.edu.in}
\and
\IEEEauthorblockN{Harichandana BSS}
\IEEEauthorblockA{\textit{Ondevice AI} \\
\textit{Samsung R\&D Institute}\\
Bengaluru, India \\
hari.ss@samsung.com}
\and
\IEEEauthorblockN{Sumit Kumar}
\IEEEauthorblockA{\textit{Ondevice AI} \\
\textit{Samsung R\&D Institute}\\
Bengaluru, India \\
sumit.kr@samsung.com}
}

\maketitle

\begin{abstract}
Emotion prediction is the field of study to understand human emotions. Existing methods focus on modalities like text, audio, facial expressions, etc., which could be private to the user. Emotion can be derived from the subject’s psychological data as well. Various approaches that employ combinations of physiological sensors for emotion recognition have been proposed. Yet, not all sensors are simple to use and handy for individuals in their daily lives. Thus, we propose a system to predict user emotion using smartwatch sensors. We design a framework to collect ground truth in real-time utilizing a mix of English and regional language-based videos to invoke emotions in participants and collect the data. Further, we modeled the problem as binary classification due to the limited dataset size and experimented with multiple machine-learning models. We also did an ablation study to understand the impact of features including Heart Rate, Accelerometer, and Gyroscope sensor data on mood. From the experimental results, Multi-Layer Perceptron has shown a maximum accuracy of 93.75 percent for pleasant-unpleasant (high/low valence classification) moods.
\end{abstract}


\section{Introduction}
One of the ways to understand user emotion is to observe and model the physiological data of the user. Physiologically and mentally, feelings regulate the state of humans. Plutchik's model of emotion, known as the "wheel of emotions" from [1], describes eight basic emotions (joy, trust, fear, surprise, sadness, disgust, anger, and anticipation). He proposed that emotions have both a positive and negative valence and that the intensity of an emotion can vary. 
Valence and arousal are two emotional characteristics that are frequently utilized in video tagging. Valence refers to an emotion's positive or negative quality, while arousal refers to the level of emotional activation. It commonly uses a number scale to express the valence and arousal levels of the emotions depicted.
James Russell developed a circumplex model and used a statistical approach to group similar emotions together in a circle [2]. 

Various approaches have been implemented for mood detection.
Among them, few record speech, facial expressions, tweet messages, human walks, and many other aspects to identify mood using machine learning [3, 4, 5].

Physiological reactions can be used to build emotion identification systems and can provide important details about a person's emotional state. Physiological sensors are tools that can gauge the body's many physiological responses, including temperature, skin conductance, heart rate, and breathing rate. These sensors can be applied to many different tasks, such as emotion identification, where they might reveal information about a person's emotional state. ElectroCardioGram (ECG) and ElectroEncephaloGram (EEG) devices may assess the electrical activity of the heart and brain, respectively. Skin conductance, which is correlated with variations in sweat gland activity, can be measured by Galvanic Skin Response (GSR) and ElectroDermal Activity (EDA) sensors. Changes in skin temperature can be detected by temperature sensors, while variations in blood flow can be detected using Blood Volume Pulse (BVP) sensors.

Although using any or a combination of the physiological sensors mentioned above has been shown to be the best option for developing highly accurate emotion recognition models, the sole goal of this paper is to understand the feasibility of emotion prediction from daily use devices like smartwatches non-intrusively. In this paper, we attempt to identify mood from heart rate, accelerometer, and gyroscope sensor data. These sensors can be used to create models with lower computational complexity and battery consumption and can enable real-time mood detection. 

A heart rate sensor often entails the use of a device, in this case, a smartwatch, that is equipped with a sensor capable of measuring the electrical activity of the heart. This sensor namely the Photoplethysmography (PPG) sensor, detects changes in blood volume in the capillaries of the wrist and uses this data to compute the heart rate. Data collection utilizing an accelerometer and gyroscope sensor often entails using a device equipped with these sensors, such as a smartphone, tablet, or wearable device. The accelerometer and gyroscope both measure acceleration and tilt, whereas the gyroscope also measures angular velocity. The sensors' data can be used to track physical activity.

Additionally, since all three of these sensors are readily accessible through any edge device, such as a smartwatch or fitness tracker, adding mood detection functionality to such a device can also assist users in addressing mental health issues at an early stage. The aim of this work is to deploy an application that accurately predicts the emotion of the person wearing it. Since there is no publicly available dataset with all the sensor readings of the simple smartwatch, data was collected from volunteers and utilized. 

In this work, we have used a Samsung smartwatch (Galaxy Watch 4) to capture data from volunteers. The watch is capable of capturing heart rate data, accelerometer, and gyroscope readings. The data is captured while showing the volunteers a set of videos from the FilmStim [6] dataset in an attempt to trigger a particular emotion in them. During the initial round of data collection, it was noticed
that some of the clips were already been seen by the volunteers and that films acquired directly from the open-source dataset were not evoking the desired emotions. Therefore a video dataset was generated with a collection
of different regional and English clips from YouTube and
FilmStim [6], which could elicit eight of Robert Plutchik’s key
emotion dimensions, based on the demographic. A total of 78 volunteer data has been collected. For data collection and mood prediction, two different application pipelines have been developed and deployed on the watch.

The paper is organized as follows: The literature review of earlier relevant publications is found in Section II. The data collection process is described in Section III. The creation of datasets and binary categorization are covered in Section IV. The main findings from the data analysis are outlined in Section V. The methods for creating and testing models are described in Section VI. The outcomes of various models and inferences are displayed in Section VII. In Section VIII, the model deployment is explained. The study is concluded in Section IX, while Section X proposes future work.

\section{Related Works}
Recently, a number of works on emotional intelligence have been put into practice. Some of these varied efforts use inputs including text[4], voice[7], facial expressions[8], and other characteristics to identify emotions[5]. 
While the use of speech and text as input has been found to produce approaches that are more accurate than those that rely on facial expressions[8], many studies have proposed a physiological sensors-based approach for mood recognition. 

Quiroz, Juan Carlos et al. utilized smartwatches in their work[9] to identify emotion from their walk and were able to achieve a maximum accuracy of 78\% in the binary classification of happiness and sadness. 
HealthyOffice[10] uses the Toshiba SilmeeTM Bar Type, a prototype wearable sensor, to collect physiological information and predict employee mood. Additionally, it provides a smartphone app and a method for gathering ground truth, creates different mood models for customized and generalized mood detection, and achieves an average classification accuracy of 70.6\% across 8 moods and 5 categories. Work done by Shu, L.; Yu, Y. et al.[11] utilizes only heart rate variable data where the mood was attempted to classify into happy, sad, and neutral and was able to achieve an accuracy of 70.4\% and 52\% for binary and five class classifications respectively. EmoSense[12] presents an ECG signal-based technique and builds a typical machine learning pipeline for 9-class emotion classification with 92.5\% accuracy using the DREAMER[23] dataset. In a study by Merve Erknay zdemir et al. [13], ANN is used to classify high/low valence with a true positive rate of 69\%.

This work differs from other works in terms of the dataset used to trigger volunteers' emotions and the sensor data collected. Sensor data such as gyroscope reading is unique. The way the data was pre-processed for modeling is different. Also, considering the battery life, which is a very important factor for devices like a smartwatch, our solution does not require continuous data collection and can predict emotion whenever heart rate is measured by the user.

\section{Data Collection}

\begin{figure*}[htbp]
\centerline{\includegraphics[scale=0.49]{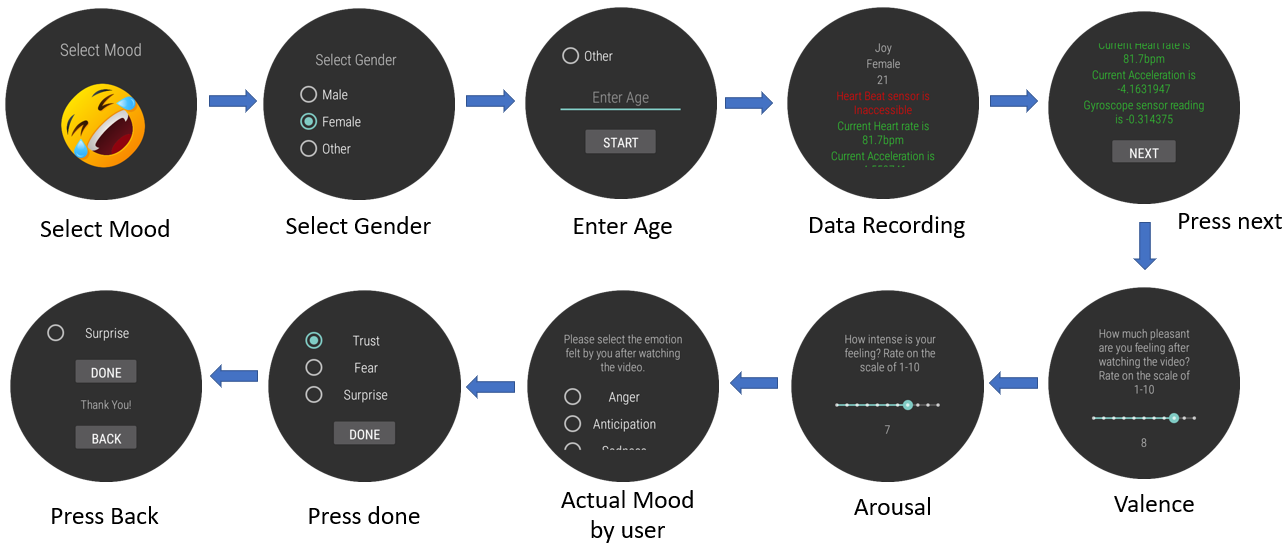}}
\caption{Data Collection App flow diagram}
\label{fig-4}
\end{figure*}

The data is collected using a Samsung Galaxy Watch across different classes of volunteers. An android wearable application with the flow mentioned in Fig. 1 has been developed and deployed on the watch to collect the user's age, gender,  heart rate, accelerometer, and gyroscope sensor data, and the emotions are evaluated through videos.

\begin{itemize}
    \item Initially the volunteers were informed about the process and were given some time to be comfortable with the system. 
    \item The volunteer was made aware that his/her data was being collected via sensors. 
    \item The emotion was initially selected by the team (which the volunteer is not aware of), and the gender and age is entered by the volunteer. 
    \item Moving ahead, the data collection process is initiated, and all the accessible sensors start working except for the heart rate sensor. The heart rate sensor takes around 15-20s to start. 
    \item Once the heart rate starts off, the video is displayed and sensor data is stored in the database in real-time simultaneously.
\end{itemize}
Post this, the user takes three self-assessments:
\begin{itemize}
    \item "How much pleasant are you feeling after watching the video? Rate on a scale of 0-10". This gives us valence value.
    \item "How intense is your feeling? Rate on a scale of 0-10". This gives us arousal value.
    \item "Please select the emotion felt by you after watching the video". This gives us the actual mood felt by the volunteer.
\end{itemize}
The valence arousal scale's purpose is to estimate the sense of pleasure and displeasure which can be used for justification of the mood [14].

Data from volunteers ranging from various age groups of 16-30, 31-45, and then above 45 for each gender were collected. The main aim for collecting the data from such a wide range of age groups was to primarily know the scale at which different emotions vary in terms of age.
A total of 78 volunteer data has been recorded.

\section{Data Processing}
\subsection{Data Sets formation}

Taking motivation from the paper [15], the collected data was further formatted into statistical data with more features resulting in two types of datasets for ML model building, non-statistical and statistical.
The non-statistical dataset has the sensor readings of that particular volunteer at a particular instance.
The statistical dataset has the statistical data of heart rate, accelerometer, gyroscope meter, and other columns like age and gender.
Along with statistical data such as mean, mode, median, and number of peaks recorded, heart rate variable metrics [16] were also determined and utilized. The following metrics were considered alongside statistical data:
\begin{itemize}
    \item SDNN, the standard deviation of time intervals between peaks(cleaned data).
    \item RMSSD, root mean square of successive time interval difference between the peaks(raw data).
    \item NN50, the total number of time intervals between peaks
that differ by more than 50ms
    \item pNN50, percentage of time intervals between peaks that differ by more than 50ms.
    \item HR-range, difference between maximum and minimum heart rate readings.
\end{itemize}

\subsection{Binary Classification - Pleasant vs Unpleasant}

Since only 78 people's data was collected and used, it was not enough for the classification of 8 moods from Robert Plutchik's Wheel of Emotions.
To deal with this, the data tagged with anger, sadness, disgust, and fear i.e. low valence moods, and the one tagged with joy, surprise, anticipation, and trust i.e. high valence moods was segregated as unpleasant and pleasant moods respectively[4].

\section{Insights from Data analysis}
The 16–30 age group has recorded the greatest average accelerometer, gyroscope, and heart rate readings of any age group, it has been noted. Males recorded the greatest average accelerometer and gyroscope readings, while females recorded the highest average heart rate.

The volunteer's mood was compared with the averages of the accelerometer, gyroscope and heart rate readings, and the following key insights were found:
\begin{itemize}
    \item "Sadness" and "Surprise" showed the highest and lowest accelerometer average respectively.
    \item "Anticipation" and "Trust" showed the highest and lowest gyroscope average respectively.
    \item "Joy" and "Trust" showed the highest and lowest mean heart rate.
\end{itemize}

 The overall unpleasant mood category was observed at the higher end of mean accelerometer and gyroscope readings while the pleasant category was observed at the higher end of mean heart rate readings. 

Post this, every data point on the valence-arousal graph in the left part of Fig. 2 was labeled with its corresponding mood type category. Two distinct clusters can be roughly distinguished.

\begin{figure}[htbp]
\centerline{\includegraphics[scale=0.24]{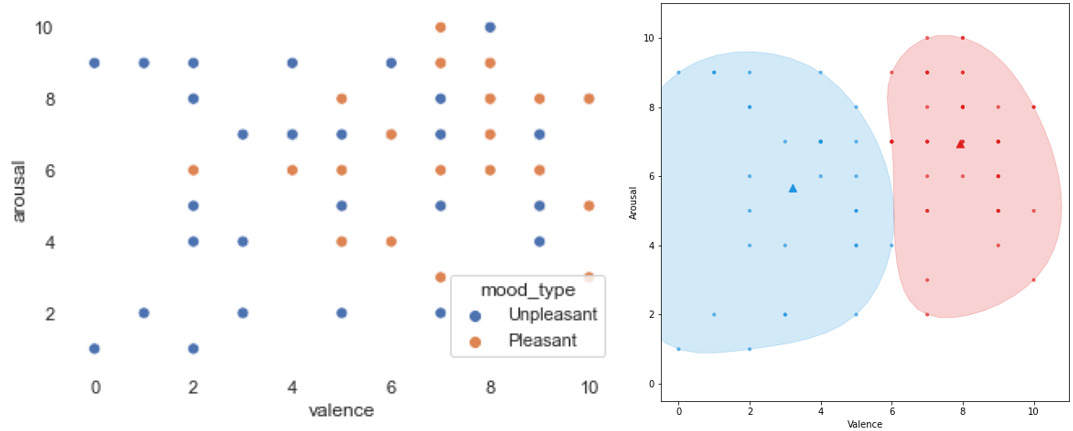}}
\caption{True values of valence and arousal and Clusters identified by K-means algorithm where k=2}
\label{fig-5}
\end{figure}

Applying the K-means algorithm, with k = 2, two well-defined clusters were identified (illustrated in the right part of Fig. 2).
Comparing the left part of Fig. 2 and the right part of Fig. 2, the algorithm was able to identify two clusters, which were able to represent an unpleasant and pleasant mood, respectively.
Thus justifying the use of valence and arousal scales for categorizing moods.

\section{ML Model building and testing}
\subsection{Train-Test Data splitting}
In order to avoid data imbalance, train-test splits of the data were made with an equal proportion of each target class in both the training and testing data.
\subsection{Models}
Logistic Regression, Decision Tree Classifier, Random Forest Classifier, Gaussian Naïve Bayes, K Nearest Neighbors, AdaBoost Classifier, Multi-Layer Perceptron(MLP), XGBoost for the non-statistical dataset, and same models in addition to SVM and Gradient Boost for statistical dataset were trained with different sets of input features after normalizing the dataset.
\subsection{Custom Accuracy function for Non-Statistical dataset}
Since the models using non-statistical data were trained to predict mood at every instance of sensor reading for a particular ID (volunteer), it was identified that a conventional accuracy function was not suitable to test these models. So, a custom accuracy function was built with a naive approach. The logic was that the mood which was predicted most of the time was taken as the final predicted mood.

\section{Results and Inference}

\subsection{Using non-statistical dataset}

\begin{table}[h!]
\caption{Binary Classification Results using Non-Statistical Dataset}
\centering
\scalebox{0.75}{%
\begin{tabular}{|c|c|c|c|c|c|c|}
\hline
{\color[HTML]{000000} \textbf{Input Features}} & {\color[HTML]{000000} \textbf{\begin{tabular}[c]{@{}c@{}}Logistic \\ Regression\end{tabular}}} & {\color[HTML]{000000} \textbf{\begin{tabular}[c]{@{}c@{}}Decision\\ Tree\end{tabular}}} & {\color[HTML]{000000} \textbf{\begin{tabular}[c]{@{}c@{}}Random\\ Forest\end{tabular}}} & {\color[HTML]{000000} \textbf{\begin{tabular}[c]{@{}c@{}}Gaussian\\ NB\end{tabular}}} & {\color[HTML]{000000} \textbf{KNN}} & {\color[HTML]{000000} \textbf{AdaBoost}} \\ \hline
{\color[HTML]{000000} \textbf{Acc, Gyro, Hr}} & {\color[HTML]{000000} 46} & {\color[HTML]{000000} 35} & {\color[HTML]{000000} 25} & {\color[HTML]{000000} 40} & {\color[HTML]{000000} 30} & {\color[HTML]{000000} 45} \\ \hline
{\color[HTML]{000000} \textbf{\begin{tabular}[c]{@{}c@{}}Acc, Gyro, Hr\\ (without age)\end{tabular}}} & {\color[HTML]{000000} 44.5} & {\color[HTML]{000000} 50.5} & {\color[HTML]{000000} 49.5} & {\color[HTML]{000000} 54} & {\color[HTML]{000000} 50} & {\color[HTML]{000000} 47.5} \\ \hline
\textbf{\begin{tabular}[c]{@{}c@{}}Acc, Gyro, Hr\\ (without gender)\end{tabular}} & 50 & 50.5 & 47 & 47.5 & 55 & 45.5 \\ \hline
\rowcolor{LightCyan}
\textbf{\begin{tabular}[c]{@{}c@{}}Acc, Gryo, Hr\\ (without age \\\& gender)\end{tabular}} & 48 & 50.5 & 52.5 & 51.5 & \textbf{56.5} & 48.5 \\ \hline
\textbf{\begin{tabular}[c]{@{}c@{}}Acc, Gyro\\ (without age \\\& gender)\end{tabular}} & 52 & 40 & 44 & 51 & 41.5 & 43 \\ \hline
\textbf{\begin{tabular}[c]{@{}c@{}}Hr\\ (without age \\\& gender)\end{tabular}} & 44.5 & 52.5 & 52.5 & 52.5 & 53.5 & 52.5 \\ \hline
\textbf{\begin{tabular}[c]{@{}c@{}}PCA\\ (3   components)\end{tabular}} & 46.5 & 42.5 & 42 & 51.5 & 44.5 & 48 \\ \hline
\end{tabular}}

\label{tab:my-table}
\end{table}

Table I shows the performance of the binary classifiers trained and tested on different sets of input features.
The average accuracies of 44\% and 46.5\% were observed with MLP and XGBoost respectively.
When using only Accelerometer, Gyroscope and Heart rate as a feature, KNN with n value = 5 showed the highest average accuracy of 56.5\%.

\subsection{Using statistical dataset}



\begin{table*}[h!]
\caption{Binary Classification Results using Statistical Dataset}
\begin{tabularx}{\textwidth}{@{} l *{9}{C} c @{}}
\toprule
\textbf{INPUT   FEATURES} & \textbf{Logistic Regression} & \textbf{Decision Tree} & \textbf{Random Forest} & \textbf{Gaussian NB} & \textbf{SVM} & \textbf{KNN} & \textbf{Gradient Boost} & \textbf{Ada Boost} & \textbf{XGB} \\ \hline
\textbf{Hrv, Hr ,Acc, Gyro} & 55 & 64.375 & 67.5 & 55 & 63.125 & 58.75 & 63.74 & 61.875 & 67.5 \\ \hline
\textbf{Hr, Acc, Gyro} & 64.375 & 60.624 & 68.125 & 60.624 & 62.5 & 53.125 & 63.125 & 65.0 & 65.0 \\ \hline
\textbf{Hrv, Acc, Gyro} & 58.75 & 58.75 & 61.25 & 56.25 & 61.25 & 60.624 & 61.25 & 57.49 & 59.375 \\ \hline
\textbf{Hrv, Hr, Acc} & 57.49 & 59.375 & 62.5 & 56.25 & 56.25 & 47.5 & 63.125 & 65.0 & 63.74 \\ \hline
\rowcolor{LightCyan}
\textbf{HRV, Hr, Gyro} & 60.624 & 66.875 & \textbf{70.625} & 60.624 & 64.375 & 63.74 & 65.0 & 62.5 & 66.875 \\ \hline
\textbf{Hrv, Hr} & 58.75 & 60.624 & 62.5 & 56.875 & 54.374 & 53.75 & 63.125 & 60.624 & 65 \\ \hline
\textbf{Hrv} & 52.5 & 46.25 & 45.0 & 58.125 & 41.25 & 49.375 & 45.0 & 46.25 & 40 \\ \hline
\textbf{Hr} & 61.875 & 60.0 & 66.25 & 59.375 & 59.375 & 63.749 & 62.5 & 57.49 & 68.125 \\ \hline
\textbf{\begin{tabular}[c]{@{}c@{}}Hrv, Hr,  Acc, Gyro \\ (without age)\end{tabular}} & 58.75 & 60.624 & 63.749 & 55 & 63.74 & 62.5 & 63.125 & 60.0 & 66.25 \\ \hline
\textbf{\begin{tabular}[c]{@{}c@{}}Hrv, Hr, Acc, Gyro \\ (without gender)\end{tabular}} & 58.75 & 62.5 & 65.0 & 55 & 63 & 55 & 59.375 & 58. 1 & 65.0 \\ \hline
\textbf{\begin{tabular}[c]{@{}c@{}}Hrv, Hr, Acc, Gyro \\ (without age \& gender)\end{tabular}} & 60.0 & 61.25 & 65.625 & 55 & 62.5 & 56.25 & 60.0 & 60.0 & 65.0 \\ \hline
\textbf{\begin{tabular}[c]{@{}c@{}}Hrv, Hr, Acc, Gyro \\ (without median \& mode)\end{tabular}} & 58.75 & 61.875 & 67.5 & 57.49 & 60.624 & 64.375 & 62.5 & 59.375 & 64.375 \\ \hline
\textbf{\begin{tabular}[c]{@{}c@{}}Hrv, Hr,  Acc,  Gyro \\ (without age,  gender, \\ median \& mode)\end{tabular}} & 60.0 & 65.0 & 68.125 & 56.875 & 64.375 & 65.0 & 61.875 & 61.875 & 65.0 \\ \hline
\textbf{Acc, Gyro} & 63.74 & 58.75 & 65.0 & 57.49 & 63.74 & 60.0 & 56.875 & 56.875 & 63.74 \\ \hline
\textbf{Acc, Gyro (without gender)} & 66.875 & 61.25 & 66.25 & 58.12 & 64.375 & 58.75 & 61.25 & 55.625 & 63.125 \\ \hline
\textbf{Acc, Gyro (without age)} & 66.875 & 59.37 & 66.25 & 56.875 & 63.125 & 60.624 & 58.125 & 58.125 & 61.25 \\ \hline
\textbf{Acc, Gyro (without age \& gender)} & 66.875 & 60.0 & 65.625 & 57.49 & 68.125 & 60.624 & 60 & 57.49 & 60.624 \\ \hline
\end{tabularx}

\label{tab:my-table}
\end{table*}

 From all the binary classifiers trained and tested using statistical data, the highest accuracy performance was recorded by Random Forest as shown in Table II. An accuracy of 70.625\% was achieved when all statistical data, heart rate variables while Gyroscope reading dropped, were given as input features.

\subsection{Using MLP with Statistical Data}

MLP is a small neural network with fully connected dense layers.
Fig. 3 shows the architecture of our MLP model, it had an input layer with 32 units followed by a dropout layer, whose rate is 0.5 followed by a fully connected dense layer of 8 units and finally an output layer.

\begin{figure}[htbp]
\centerline{\includegraphics[scale=0.4
]{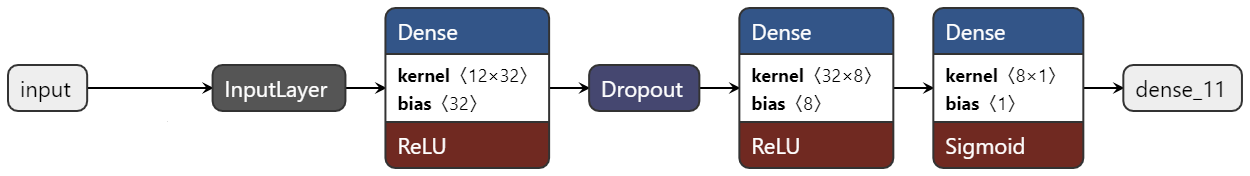}}
\caption{The MLP model summary.
}
\label{fig-7}
\end{figure}

Similar to the Machine Learning model approach in the previous section, the MLP model was tested with various input features. The highest accuracy achieved was 93.75\% while the input features were statistical data of accelerometer and gyroscope meter readings while gender was dropped. Fig. 4 and Fig. 5 indicate a decrease in loss and an increase in accuracy respectively with an increase in epochs.

While making sure either Heart Rate data and Heart Rate Variable data or both were present, the highest accuracy recorded was 87.5\% while using 1) all statistical data but dropping age and gender and 2) all statistical data but dropping age alone were input features. The results with other sets of input features are mentioned in Table III.

\begin{table}[htbp]
\caption{Performance of MLP}
\centering
\scalebox{0.9}{%
\begin{tabular}{|c|c|c|}
\hline
\textbf{INPUT   FEATURES} & \textbf{Training Accuracy} & \textbf{Testing Accuracy} \\ \hline
\textbf{Hrv, Hr ,Acc, Gyro} & 96.77 & 68.75 \\ \hline
\textbf{Hr, Acc, Gyro} & 98.38 & 68.75 \\ \hline
\textbf{Hrv, Acc, Gyro} & 93.54 & 75 \\ \hline
\textbf{Hrv, Hr, Acc} & 88.70 & 56.25 \\ \hline
\textbf{HRV, Hr, Gyro} & 93.54 & 81.25 \\ \hline
\textbf{Hrv, Hr} & 87.09 & 56.25 \\ \hline
\textbf{Hrv} & 77.41 & 31.25 \\ \hline
\textbf{Hr} & 82.25 & 56.25 \\ \hline
\textbf{\begin{tabular}[c]{@{}c@{}}Hrv, Hr,  Acc, Gyro \\ (without age)\end{tabular}} & 91.93 & 87.5 \\ \hline
\textbf{\begin{tabular}[c]{@{}c@{}}Hrv, Hr, Acc, Gyro \\ (without gender)\end{tabular}} & 93.54 & 81.25 \\ \hline
\textbf{\begin{tabular}[c]{@{}c@{}}Hrv, Hr, Acc, Gyro \\ (without age \& gender)\end{tabular}} & 96.77 & 87.5 \\ \hline
\textbf{\begin{tabular}[c]{@{}c@{}}Hrv, Hr, Acc, Gyro \\ (without median \& mode)\end{tabular}} & 96.77 & 81.25 \\ \hline
\textbf{\begin{tabular}[c]{@{}c@{}}Hrv, Hr,  Acc,  Gyro \\ (without age,  gender, \\ median \& mode)\end{tabular}} & 91.93 & 62.5 \\ \hline
\textbf{Acc, Gyro} & 90.32 & 81.25 \\ \hline
\rowcolor{LightCyan}
\textbf{Acc, Gyro (without gender)} & \textbf{85.48} & \textbf{93.75} \\ \hline
\textbf{Acc, Gyro (without age)} & 87.09 & 87.5 \\ \hline
\textbf{Acc, Gyro (without age \& gender)} & 85.48 & 87.5 \\ \hline
\end{tabular}}
\label{tab:my-table}
\end{table}

\begin{figure}[htbp]
\centerline{\includegraphics[scale=0.45]{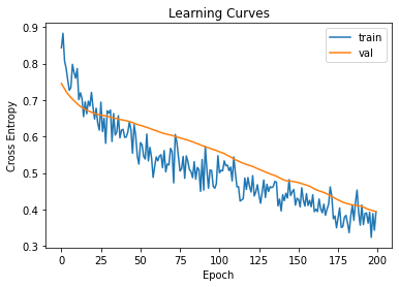}}
\caption{The training and testing losses are 
plotted against the epochs.
}
\label{fig-8}
\end{figure}

\begin{figure}[htbp]
\centerline{\includegraphics[scale=0.45]{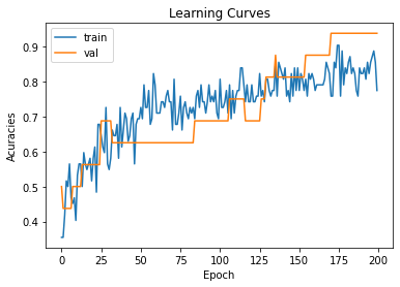}}
\caption{The training and testing accuracies are 
plotted against the epochs.
}
\label{fig-9}
\end{figure}

\begin{figure}[htbp]
\centerline{\includegraphics[scale=0.4]{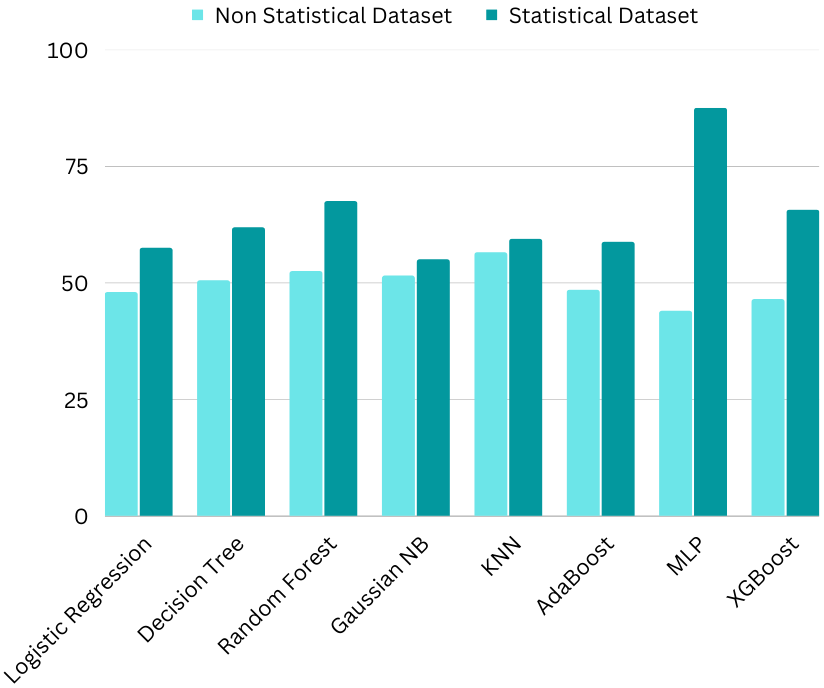}}
\caption{Bar graph representation of model performance on both datasets.}
\label{fig-10}
\end{figure}
As depicted in Fig. 6, higher accuracy was observed for each model using the statistical dataset as compared to that of the non-statistical dataset.
EEG and ECG data have been found to be helpful for identifying emotions[12]. Though these physiological sensors were not employed, our work performed better than a few others who did. While some works aim to categorize all opposing emotions in a binary fashion, others tend to do it in a way that is similar to ours [13, 17, 18, 19].

\section{Deployment}
After all the models were successfully studied, a complete TensorFlow pipeline was built, incorporated into a different application that was identical to the one discussed earlier, and it was then deployed on the Samsung Galaxy watch. When the user initiates the app, it takes user input and records sensor data to determine the user's mood type and then displays it.

\section{Conclusion}
As proposed, our work was able to predict the user's mood as pleasant or unpleasant for a specific duration without any use of ECG, EEG, GSR, EDA, BVP, etc., sensors data. Nine Supervised Machine Learning models and a neural network were trained on two different dataset formats and the results have been compared.
Taking into consideration of being restricted to gathering and using heart rate, gyroscope, and accelerometer data through a single smartwatch on our own, our work has achieved a maximum accuracy of 93.75\% with MLP and outperformed binary classifiers from few other works which are either trained on readily available EEG data and multi peripheral physiological signals or on their own physiological signals data.

\section{Future Works}
In this work, we demonstrated that with daily use devices like smartwatches, it is possible to detect user emotion non-intrusively with just the use of existing sensors. The proposed machine learning models classify data into pleasant and unpleasant categories using binary classification, with four pleasant emotions (joy, anticipation, surprise, trust) and four unpleasant emotions (anger, sadness, disgust, and fear). In the future, an extension to the model can be developed that further classifies each category into individual emotions. 
The majority of emotion classification or recognition makes use of EEG and ECG data. Our current model has been built on heart rate, accelerometer, and gyroscope sensor data and still achieves comparable performance.
The Deep Neural Network(DNN) model, can be researched further for understanding more accurate patterns and identifying the primary emotions. This can be supported with an extended data set along with EEG and ECG data under medical supervision to improve the classification accuracy.

\vspace{12pt}

\end{document}